\documentclass[9pt,twocolumn,twoside]{gsajnl}

\usepackage{epstopdf}
\usepackage{float}
\usepackage{textgreek}
\usepackage[section]{placeins}
\usepackage{gensymb}
\usepackage{siunitx}
\articletype{inv} 

\runningtitle{Systematic review} 
\runningauthor{Jittprasong}

\title{Mesenchymal stem cells as carrier cells to enable effective intratumoral delivery of oncolytic virus for oncolytic virotherapy: a systematic review}

\author[1,$\ast$]{Chottiwatt Jittprasong} 

\affil[1]{Department of Biomedical Engineering, College of Engineering, City University of Hong Kong}

\correspondingauthoraffiliation[$\ast$]{Corresponding author: Department of Biomedical Engineering, College of Engineering, City University of Hong Kong, Kowloon, Hong Kong. \href{chottiwatt.j@my.cityu.edu.hk}{chottiwatt.j@my.cityu.edu.hk}}

\begin{abstract}
Oncolytic viruses, which may be naturally occurring or genetically engineered, are a type of virus that infects and destroy cancer cells preferentially. Owing to their selectivity, they outperform conventional chemotherapy and radiotherapy, which both have a tendency to impact non-target cells and cause unwanted adverse side effects. Oncolytic virotherapy is a type of cancer treatment in which oncolytic viruses are deliberately introduced into patients affected with cancers in order for them to infect and destroy cancer cells locally or systemically, in a manner analogous to chemotherapy but with a greater degree of selectivity. Multiple studies indicate that oncolytic virotherapy is effective in vitro but in vivo findings remain ambiguous due to the approach's primary limitation: inefficient therapeutic agent delivery to its target, which is heavily influenced by the immune system. Here, we propose overcoming this limitation by exploiting a recent discovery in cancer research: a carrier cell. By exploiting their tumor-promoting activities, mesenchymal stem cells may be employed for cancer therapy by serving as a carrier for the oncolytic viruses toward their target. This approach directly addresses the limitation of conventional oncolytic virotherapy, where oncolytic viruses are often poorly delivered after systemic administration. 
\end{abstract}


\dates{\rec{28 03, 2022} \acc{28 03, 2022}}

\begin{document}

\maketitle
\thispagestyle{firststyle}
\vspace{-13pt}

\section{Introduction}

Cancer is a group of diseases that affect a large number of humans and animals. Such diseases are characterized by abnormal cell growth, invasion, metastasis, and organ dysfunction and may ultimately result in death. Humans have over 100 different forms of cancer, which are classified both clinically and histologically \citep{nci_2007}. While some cancers respond to conventional therapy or exhibit transient remissions, others have been categorized as aggressive and difficult, if not impossible, to cure with traditional therapies, such as glioblastoma multiforme (GBM), the most aggressive form of brain cancer \citep{RN49}. In 2015, cancer affected over 90 million people globally and claimed over 9 million lives \citep{RN50, RN51}. Efforts to cure cancer have been ongoing for decades. Several treatment protocols have been established and proven effective for the treatment of many types of cancer.

Chemotherapy and radiotherapy are two of the most frequently used therapies for a wide variety of tumors. They are undeniably successful therapies that have cured millions of people. Additionally, chemotherapy and radiotherapy have become substantially more efficient as a result of continual research. Despite these accomplishments, both therapies depend on the use of poorly selective, highly toxic substances that are administered locally or systemically and eventually have deleterious effects on adjacent normal cells. This invariably leads to dose-limiting toxicity, which discourages patients from adopting the therapy. Additionally, both therapies are linked with considerable physical and psychological side effects that may have a negative influence on patients' and families' quality of life.

\begin{figure}
\includegraphics[width=\linewidth]{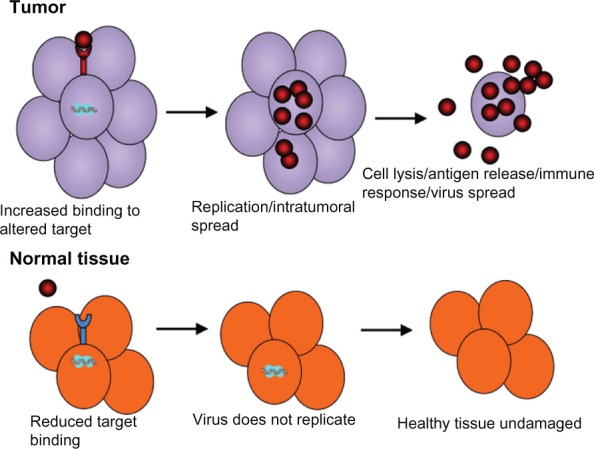}
\caption{OVs work by infecting tumor cells but sparing health cells.\\
Note: From "Oncolytic virotherapy: the questions and the promise," by \cite{RN74}, \textit{Oncolytic Virother}, 2, 19-29. Copyright 2013 by Dove Medical Press Limited under CC BY-NC 3.0 license. Reprinted under license terms.}
\label{fig:intendedinfection}
\end{figure}

Oncolytic virotherapy (OVT) is a revolutionary technique in cancer treatment that utilizes an oncolytic virus (OV) that infects and destroys cancer cells preferentially \citep{RN52}. By using nature's evolutionary process, this method leverages viruses as live therapeutic agents. In fact, OVs are quite naturally occurring in our environment and have an undeniable track record against cancer, such as senecavirus \citep{roberts_2006_naturally}. OVs are able to infect and replicate in cancer cells before destroying these cells \hyperref[fig:intendedinfection]{(Figure 1)}. Genetic engineering techniques have been used to modify OVs for increased tumor selectivity, higher replication rate, and more toxicity toward cancer cells to enhance therapy effectiveness while minimizing off-target effects \citep{RN53}. OVs' inherent genetic flexibility enables them to be efficient against a diverse variety of malignancies with varying genetic origins.

OVT provides various advantages over conventional cancer therapy, but it also has substantial drawbacks in terms of safety and efficacy. Since OVT is an intentional inoculation of potentially virulent viruses, its safety is considerable. Utilizing OVT requires close monitoring to ensure safe and effective levels are maintained at the target location. Efficacy of OVT is also affected by an inability to evade the adaptive immune system, inefficient delivery, and infection of non-target cells. Therefore, OVT requires a development to produce effective mechanisms of action to overcome these limitations.

\begin{table}
\begin{tableminipage}{\linewidth}
\caption{OVs under clinical trial \citep{RN74}}
\begin{tabularx}{\linewidth}{c c c}
\hline
{\bf Type of virus} & {\bf Clinical trial}\footnote{completed and ongoing.}& {\bf Reference}  \\
\hline
Adenovirus &  Phase I*& \cite{RN76}\\
Vaccinia virus  & Phase II*&\cite{RN77}\\
Herpes Simplex Virus & Phase II*&\cite{RN78}\\
\hline
\end{tabularx}
\end{tableminipage}
\label{table1}
\end{table}

In this literature review paper, I will discuss the limitations of OVT and the bioengineering approach to overcome these limitations.

\section{Discussion}
\label{sec:discussion}

The primary constraint of contemporary OVT is the OV's ability to reach the targeted cancer cells. Several factors contribute to inefficient delivery, including the following:

\begin{itemize}
\item Immune system
\item Non-cancer cells infection
\end{itemize}

In case of successful and sufficient delivery of OVs, the effectiveness of OVT remains contingent on a number of parameters, including but not limited to the following \citep{GORADEL2021100639}: 

\begin{itemize}
\item Tumor penetration
\item Hypoxic effect
\end{itemize}

\subsection{Immune System}
Administration of OVs as therapeutic agents can be done locally or systemically. In a local administration, OVs are delivered intratumorally. This type of administration is restricted to easily accessible cancer such as melanoma. At the same time, inaccessible tumors will need systemic administration through intravenous (IV) injection.

\begin{figure}
\includegraphics[width=\linewidth]{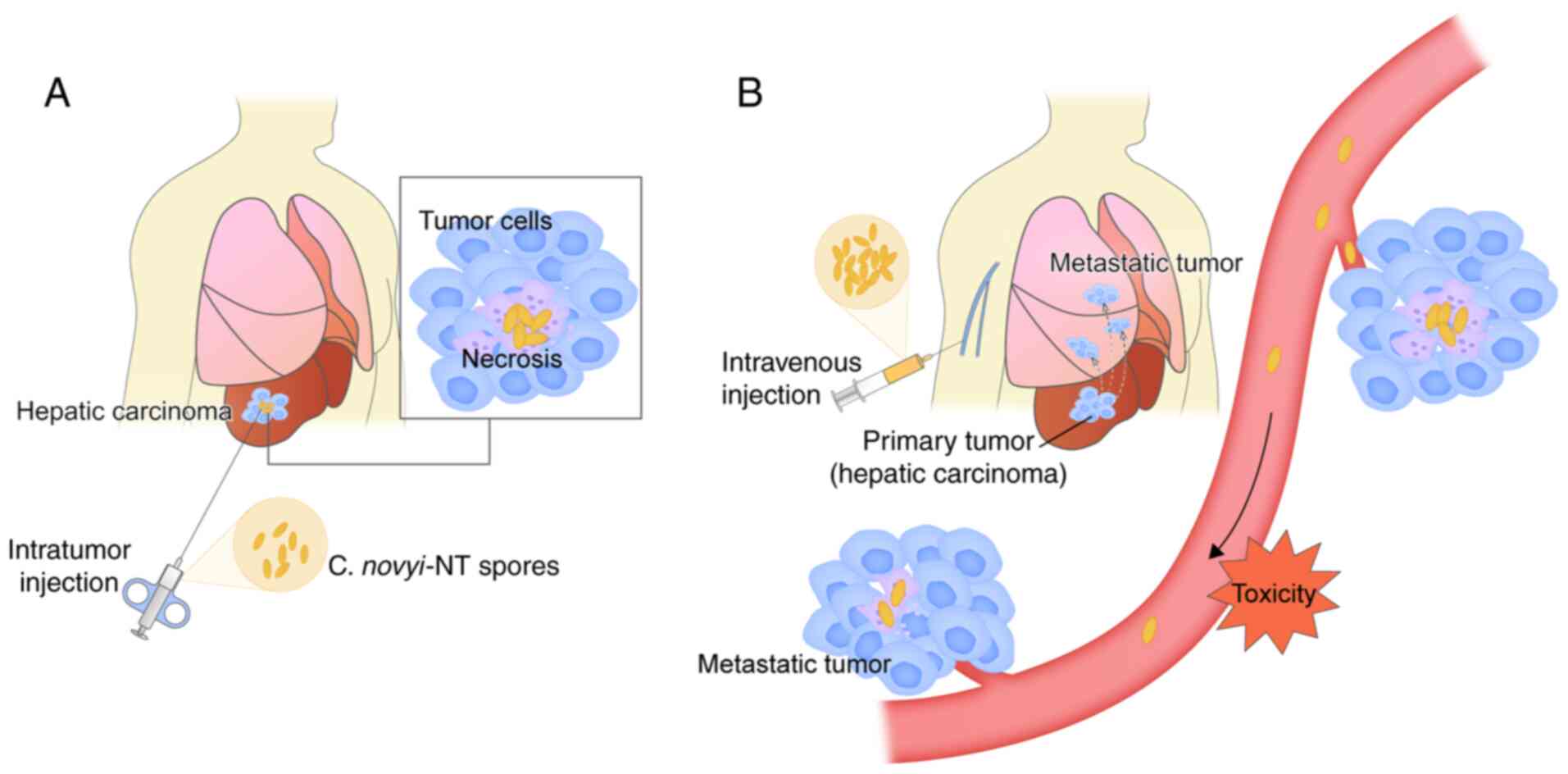}
\caption{\textbf{(A)} Intratumoral injection for hepatic cancer. \textbf{(B)} Systemic injection for metastasized cancers\\
Note: From "Novel insights into the role of Clostridium novyi‑NT related combination bacteriolytic therapy in solid tumors (Review)," by \cite{RN65}, \textit{Oncology Letters}, 21(2), 110. Copyright 2020 by Feng et al. under CC BY-NC-ND 4.0 license. Reprinted under license terms.}
\label{fig:2}
\end{figure}

As seen in \hyperref[fig:2]{Figure 2}, when administered intravenously, the immune system may act as a double-edged sword; the therapeutic agent must avoid being neutralized by neutralizing antibodies (Nabs), which may be from a pre-existing immunity, in the blood circulation before reaching its target \citep{NABS}. While circulating in the body, it must also avoid the subsequent immune system's natural response to infection and complement system that attempt to eliminate the therapeutic agent prematurely as well \citep{prematureNeutralization}.

Immunosuppressive medications and chemotherapy to inhibit the complement system may be administered to prolong the duration of the oncolytic infection, thus prolonging the therapeutic effect \citep{immunosupressive, immunesystemInhibition1, immunesystemInhibition2, immunesystemInhibition3}. However, the use of immunosuppressive drugs may cause toxicity \citep{immunosupressiveNeuroToxic, immunosupressiveCVToxic}. A more viable solution would be to allow the OV to bypass the immune system or even use the immune system to aid in tumor clearance since the inflammation induced by the OV infection within the tumor contribute to immune recruitment and subversion of immune evasion by the tumor cells \citep{immuneBypass1,immuneBypass2,immuneSubversion}.

\begin{figure}[H]
\includegraphics[width=\linewidth]{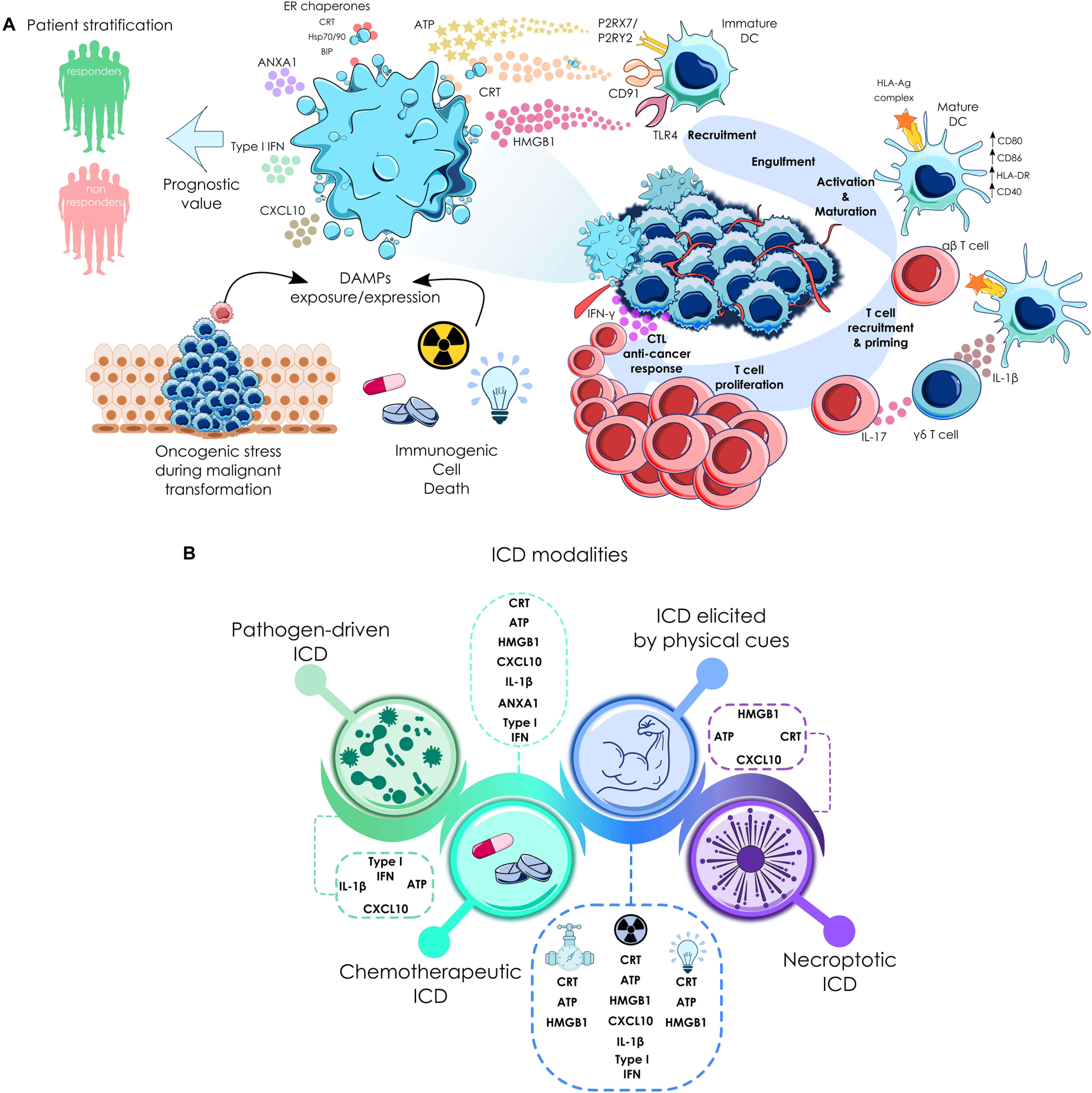}
\captionsetup{belowskip=0cm}
\caption{Immunogenic cell death cycle \textbf{(A)} Stressed cancer cells undergoing immunogenic cell death will elicit DAMPs by exposing calreticulin (CRT), endoplasmic reticulum chaperones, such as Hsp70, Hsp90 or Bip on the surface, secreting ATP, mediating type I interferon (IFN) response triggering production of CXC-chemokineligand 10 (CXCL10), and releasing high-mobility group box 1 (HMGB1) and annexin A1 (ANXA1). Binding of these extracellularly exposed molecules allows the engulfment of cell corpses by APCs, including DCs. Basically, these DAMPs function as an "eat-me" signal. These immunostimulatory signals will eventually lead to stimulation of adaptive immune response and establishment of cytotoxic T lymphocyte (CTL)-mediated antitumor response via IFN-\textgamma dependent mechanism. \textbf{(B)} Different causative factors elicit different form of signaling pathways. Pathogen-driven ICD consits of Type I IFN, Interleukin 1 Beta (IL-1\textbeta), Adenosine Triphosphate (ATP), and CXCL10.\\
Note: From "Immunogenic Cell Death and Immunotherapy of Multiple Myeloma" by \cite{10.3389/fcell.2019.00050}, \textit{Frontiers in Cell and Developmental Biology}, 7. Copyright 2019 by Serrano-del Valle, Anel, Naval and Marzo under CC BY license. Reprinted under license terms.}
\label{fig:ICD}
\end{figure}

Infection of tumor cells by OVs and its immunologic cell death (ICD) induces the local and systemic antitumor immunity mediated by the adaptive immune system, resulting in subsequent cancer remission. Danger-associated molecular patterns (DAMPs) released during ICD from cancer cells, including HMGB1, ATP, ecto-CRT, and type I IFNs, attract the antigen-presenting cells (APCs), especially dendritic cells (DCs), whereby they acquire tumor antigens (Ags), matured, and prime the T cell-mediated immune response \hyperref[fig:ICD]{(Figure 3)}. Pro-inflammatory cytokines secreted by DCs from the binding of DAMPs and tumor Ags presented to tumor-specific CD8\textsuperscript{+} T cells mediated the antitumor immunity by secretion of effector cytokine and cytolysis \citep{DAMP,10.3389/fcell.2019.00050}.

Additionally, OVs may operate as viral vectors for gene expressions suppressing cancer's angiogenesis process, so reducing tumor development by depriving it of oxygen \citep{RN66, RN67}.

It can be assumed that breaking cancer's immunological tolerance by recruiting the immune system's assistance in destroying the cancer is a vital part of OVT's success \citep{DAMP,immuneSubversion}. 

However, nature functions differently since the immune system is predisposed to clear the infection regardless of the virus's goal, studies have shown \citep{RN61, RN63}. OVs, particularly when administered systemically, should remain in the body long enough to infect and kill cancer cells while also stimulating antitumor immunity mediated by the adaptive immune system. By activating the adaptive immune system, OVs must also withstand the subsequent adaptive immune system's attempt to clear them long enough for therapeutic effects to manifest \citep{RN64}.

It is critical to discover safe methods for OVs to circumvent the immune system in order to maximize the advantages of both the immunological and OVT to achieve significant cancer therapeutic efficacy. 

\subsection{Non-cancer cells infection}
Multiple strategies have been proposed and experimented with to prevent the OVs from the non-cancer cells, including transcriptional targeting, genetic modification, chemical modification, alterations of viral genes making them incompatible for replication in normal cells, microRNA-based regulation of viral gene expression \citep{MDPI1, Frontier1}. Although these studies demonstrated that OVs preferentially infect cancer cells and significantly spare normal cells, numerous studies also suggest otherwise.

Despite the high tolerance of non-cancer cells to OV infection in clinical trials, it is believed that extensive infection of non-target cells may result in serious adverse events or impair the therapeutic efficacy under extenuating conditions \citep{NAUMENKO2022663}.

\begin{figure}[H]
\includegraphics[width=\linewidth]{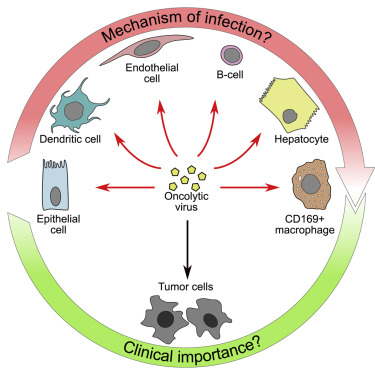}
\caption{Diagram demonstrating ability of OVs to infect numerous type of non-cancer cells.\\
Note: From "Infection of non-cancer cells: A barrier or support for oncolytic virotherapy?," by \cite{NAUMENKO2022663}, \textit{Molecular Therapy - Oncolytics}, 24, 663-682. Copyright 1969 by Elsevier under CC BY 4.0 license. Reprinted under license terms}
\label{fig:non-cancer}
\end{figure}

Various non-cancer cell infections have not yet been examined, and the mechanisms of infection are mainly unknown \hyperref[fig:non-cancer]{(Figure 4)}. The majority of studies reporting in vivo infection with OVs use a treatment protocol that involves intravenous delivery of the therapeutic drugs. To minimize widespread infection, the majority of studies concentrate on particular cancer cells or organs, such as tumor endothelial cells (ECs) or the spleen. Nonetheless, non-cancer cell infections are seen in all types of viruses and their target tissues or organs.

\FloatBarrier
\FloatBarrier

\FloatBarrier
\begin{figure}[H]
\includegraphics[width=\linewidth]{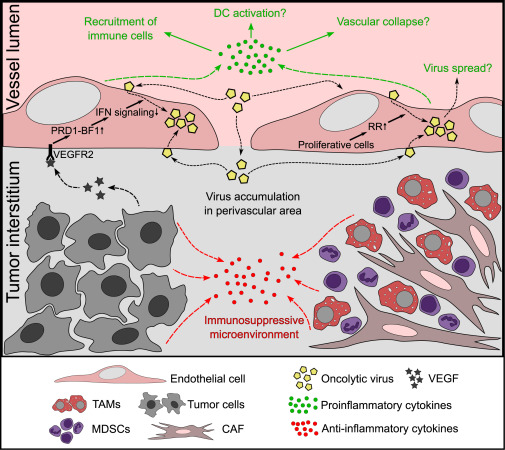}
\captionsetup{belowskip=0pt}
\caption{Numerous mechanisms contribute to ECs vulnerability to OV infection: (1) increased permeability of neovessels allows for the accumulation of OVs; (2) highly active proliferation of ECs provides OVs with an enzyme required for replication; and (3) VEGF binding to VEGFR2 activates a repressor that inhibits type I interferon-mediated antiviral signaling.\\
Note: From "Infection of non-cancer cells: A barrier or support for oncolytic virotherapy?," by \cite{NAUMENKO2022663}, \textit{Molecular Therapy - Oncolytics}, 24, 663-682. Copyright 1969 by Elsevier under CC BY 4.0 license. Reprinted under license terms.}
\label{fig5}
\end{figure}
\FloatBarrier

\subsubsection{Mechanism of infection in cancer cells: }

Tumor EC is most frequently detected for OV infection by vaccinia virus (VV), herpes simplex virus (HSV), reovirus, vesicular stomatitis virus (VSV), and adenovirus. The infection elicited by a variety of viruses across families indicates that the mechanisms of infection are due to the properties of tumor ECs rather than viral properties. The EC's susceptibility to OV infection is mostly due to vascular endothelial growth factor (VEGF), as believed by various studies. \cite{ARULANANDAM2015210} have proved that permissiveness of EC to OV infection is largely contributed by VEGF as seen in VV-treated mice with suppression of VEGF/VEGFR2 by Abs.

Cancer cells also attract leukocytes and reprogram them to generate immunosuppressive cytokines such as tumor growth factor 1 (TGF-1) and interleukin 10 (IL-10), hence promoting viral replication and favoring OVs \citep{10.3389/fonc.2019.01115}. Proliferative activity that is greater than usual and its metabolism under the influence of angiogenic factors also establish a vulnerability for OVs to infiltrate \citep{KOTTKE20111802}.

Finally, tumor neovessels have a high permeability \citep{RN68}, which allows viruses to remain in the microenvironment \hyperref[fig5]{(Figure 5).}

\subsubsection{Mechanism of infection in non-cancer cells: }

Increased concentrations of several kinds of OVs in the spleen have been seen in studies, indicating possible infection. The primary target of OVs infection in non-cancerous cells is CD169+ macrophages of marginal metallophillic macrophages (MMMs) and subcapsular sinus macrophages (SSMs). At least two viruses, VSV \citep{NAUMENKO201814,RN69} and MV \citep{doi:10.1089/104303403322495070}, were found to infect CD169+ macrophages. 

\FloatBarrier
\begin{figure}[H]
\includegraphics[width=\linewidth]{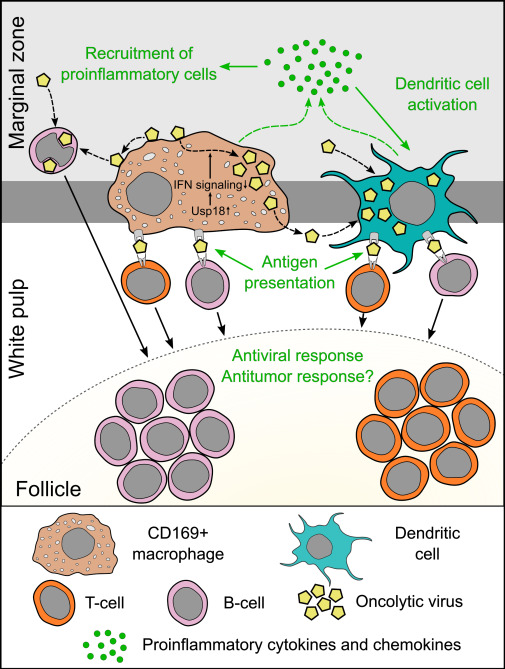}
\caption{Properties of MMMs determine permissiveness toward OVs. Anatomical location of of both MMMs and SSMs also contributed to its prevalence toward OV infection.\\
Note: From "Infection of non-cancer cells: A barrier or support for oncolytic virotherapy?," by \cite{NAUMENKO2022663}, \textit{Molecular Therapy - Oncolytics}, 24, 663-682. Copyright 1969 by Elsevier under CC BY 4.0 license. Reprinted under license terms.}
\label{fig6}
\end{figure}
\FloatBarrier

SSMs and MMMs are located in the spleen's sinusoid system, where blood slows down and allows them to search for antigens. SSMs are also present in lymph nodes, where they fulfill a similar function to antigen scanning. Despite being exposed to a variety of pathogens as part of their primary function, CD169+ cells are incapable of eliminating viruses on their own \citep{RN70}. However, the mechanism by which OVs infect CD169+ cells is unknown and is actively being explored. Due to their location and incapacity to eliminate viruses on their own, they have increased viral sequestration in their immediate vicinity \hyperref[fig6]{(Figure 6)}.

In vitro infection of DCs by OVs is also reported widely \citep{RN71, DC, Michael,ZhangLiang,ZHANG20141320}. \cite{doi:10.1128/JVI.00480-15} suggested that DCs acquire the infection from infected SSMs. Although the fate of non-cancer cells infected with OV has not been investigated, it is anticipated that non-attenuated viruses may disturb the macrophage layer. Following VV and murine cytomegalovirus (MCMV) infection, a large drop in the number of SSMs is seen, which may be due to viral lysis or IFN-mediated apoptosis. 

Wide reports of major absorption of systemically administered OVs by phagocytosis are also documented \citep{NAUMENKO201814}. Monocytes are found to be uniquely susceptible to reovirus, VSV, VV, and influenza A virus \citep{ILETT20141851, 10.1158/2326-6066.CIR-18-0309,10.1371/journal.ppat.1006914,MACLEOD2015142,RN72}. 

It is advisable to do an investigation to establish the clinical significance of these non-cancerous cell infections in order to avoid any serious adverse consequences.

\subsection{Tumor Penetration}

One of the most significant barriers to the entry of therapeutic agents into cancer cells in the presence of intracellular junctions of ECs \citep{LIPINSKI19973,10.1242/jeb.006114,Green}. Metastasis via epithelial-to-mesenchymal transition (EMT) and mesenchymal-to-epithelial transition (MET) causes the junctions to become tighter, making therapy more challenging.  \citep{10.1158/0008-5472.CAN-06-0410, RN73}. Adenoviruses are a good example of this since they are often inhibited by the existence of this barrier \citep{doi:10.1089/hum.2016.022}.

There have been a variety of ways developed to get around this issue. Oncolytic adenoviruses expressing hyaluronidase allow for enhanced cell fusion in a melanoma xenograft model as a consequence of increased lysis of the extracellular matrix \citep{RN75}. Additionally, the extracellular matrix (ECM) may impede therapeutic agent dispersal inside solid tumors \citep{ECM,https://doi.org/10.1002/glia.10204} while also retarding invasion by OVs \citep{} \citep{WOJTON2010127}. It has been shown that pretreatment of the tumor with collagenase, which breaks down the ECM layer, improves efficiency \citep{doi:10.1089/104303400750035744}.

\subsection{Hypoxic Effect}

Sustained hypoxia is known as a common condition developed in the immediate vicinity of cancer cells as a result of uncontrolled cell proliferation and an excessive amount of metabolic activity. The hypoxic effect has been demonstrated to have contradictory consequences on OVs \citep{v2010078}. \cite{RN79,RN80} have shown that adenovirus's ability to replicate and lysis has been significantly diminished as a result of the hypoxic effect. 

Modifying the E1A gene regulated by a hypoxia-response element-containing promoter expression, \cite{doi:10.1089/104303402760293574} has created an oncolytic adenovirus that can replicate effectively in a hypoxic environment.

\section{Limitations}
Systemic administration of OVs is considered the most beneficial method of conducting OVT because it influences all types of cancers and disseminated cancers, including brain tumors, the majority of which cannot be treated with radiotherapy or chemotherapy due to the inability of most therapeutic agents to cross the blood-brain barrier (BBB); however, viruses, such as semliki forest virus (SFV), can cross the BBB and infect brain cells \citep{doi:10.1126/science.424742}, this can be modified to make it preferentially target the abnormal brain cells while sparing normal cells as well. In vivo therapy with modified SFV of mouse neuroblastoma and GBM cell lines, as well as patient-derived human glioblastoma cell cultures (HGCC), demonstrated a promising effect, extending survival and curing half of the animals tested \citep{10.1158/1078-0432.CCR-16-0925}. This demonstrated the potential of OVT to even cure the disease considered incurable, GBM. 

As discussed in \nameref{sec:discussion} the current main limitations of OVT lie in circumvention of the immune system and effective delivery of systemically administered OVs. 

\section{Hypotheses}
By circulating in the bloodstream as pathogens, OVs tend to be neutralized by various natural responses of the body to clear the infection, thus, failing the OVT. 

By integrating OVs into circulating normal body cells without inflicting distress to them, OVs may utilize the normal body cells as a "carrier" to move stealthily to their target and infect it, thereby evading neutralization. Thus, effectively solving the problem of neutralization by the immune system during systemic administration.

\section{Approach}
To act as a carrier for the cancer cells, the cells must be somewhat needed by cancer itself.  Cells associated with tumors are called tumor-associated cells. Common examples of them are tumor-associated macrophages (TAM) which inhibit anti-tumor immune response, cancer-associated fibroblast (CAF), which promote remodeling of the ECM and stimulation of angiogenesis, and myeloid-derived suppressor cells (MDSC) which suppress the immune system. These cells are commonly recruited by the cancer cells to promote their growth. 

\begin{figure}[H]
\includegraphics[width=\linewidth]{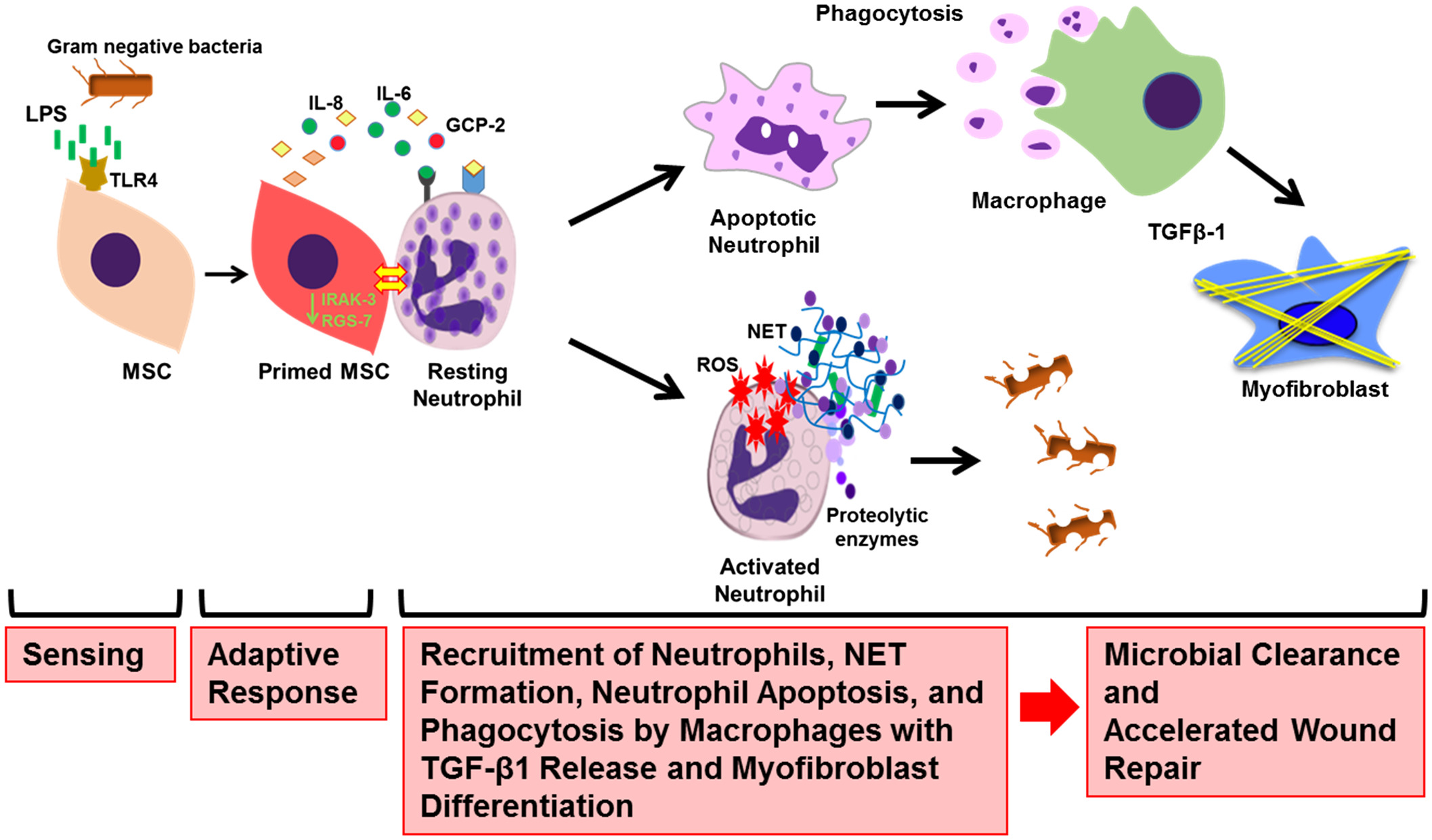}
\caption{Graphical summary showing an involvement of the MSC in wound healing process orchestrated by numerous types of cells.\\
Note: From "TLR4-dependent shaping of the wound site by MSCs accelerates wound healing," by \cite{https://doi.org/10.15252/embr.201948777}, \textit{EMBO reports}, 21(5), e48777. Copyright 2020 by John Wiley and Sons. Reprinted with permission.}
\label{fig:woundhealing}
\end{figure}

Mesenchymal stem cells (MSCs) are a kind of multipotent stem cells found throughout the body. As seen in \hyperref[fig:woundhealing]{Figure 7}, MSCs play a critical role in tissue healing by undergoing cellular differentiation to orchestrate structural repair, immunological regulation, and growth factor secretion to drive vascularization and epithelialization, as well as recruitment of resident stem cells \citep{RN81,RN82,RN83}.

Tumors may be regarded as "wounds that never heal," and in response to this prolonged wound, MSCs are continuously recruited to the tumor micro-environment to aid its reparation by secreting multiple growth factors, chemokines, and cytokines that promote the proliferation of cancer cells such as angiogenesis while also suppressing the immune response.

This way, MSCs are now tumor-associated MSCs (TA-MSCs) \citep{RN84}. But many studies also suggest that MSCs might play a role in suppressing cancer \citep{10.1371/journal.pone.0006278, 10.1002/stem.559, ijms161226215}.

\FloatBarrier
\FloatBarrier

Regardless of whether MSCs promote or repress tumor development, we can take advantage of the fact that MSCs are recruited directly to the site of cancer proliferation. By loading MSCs with OVs, we can migrate OVs directly to the site of cancer. Additionally, MSCs have been shown to protect virus that has been pre-loaded against immune-mediated neutralization \citep{10.1634/stemcells.2007-0758,10.1158/1078-0432.CCR-09-1292,10.1158/1535-7163.MCT-05-0334}. Therefore, OVs are efficiently hidden from the immune system by employing MSCs as a carrier. This way, MSCs are effectively acting as "trojan horses," and we may deliver OVs systemically without concern of the immune system neutralizing them before they reach their target. 

By combining OVs with MSCs, OVs may directly target the tumor, circumventing almost all of the constraints associated with contemporary OVT.

\section{Methods}
\subsection{Oncolytic virus}
OVs of several types may be used. ICOVIR-5, on the other hand, would be the finest fit. ICOVIR-5 is an oncolytic adenovirus whose replication is regulated by the E2F-responsive promoter. It incorporates the Arg-Gly-Asp (RGD) motif to increase infectivity and deletion of the retinoblastoma protein (pRB)-binding site to prevent it from replicating via the activated RB pathway, a pathway that can be used to differentiate normal cells from cancer cells \citep{RUANO20201033,CASCALLO20071607}.

The quantity of ICOVIR-5 replicated inside ex vivo proliferating MSCs is found to be 3-log lower than the number of cancer cells \citep{RN84}.

\subsection{Carrier mesenchymal stem cell}
The protocol is based on a work established by \cite{RN86}. Autologous MSCs can be collected from the patient through bone marrow aspirate (BMA). Isolation of mononuclear cells can be done by density gradient centrifugation (400 g, 25 min, 20 \textdegree C). After centrifugation, the substance is washed twice with phosphate buffer by sedimentation at room temperature for 5 minutes (300 g), resuspended in MSC medium of Dulbecco's Modified Eagle Medium (DMEM), 10\% Fetal Calf Serum (FCS), 2mM glutamine, 10 U/L penicillin, and 10 \micro g ml\textsuperscript{-1} streptomycin added to the culture at 1.5 x 10\textsuperscript{5} cells per cm\textsuperscript{2} and left to adhere for 24 hours. 

Non-adherent hematopoietic stem cells can be removed by washing twice with phosphate-buffered saline (PBS) at 10ml. MSC can then be cultured at 37 \textdegree C at 5\% CO\textsubscript 2 in an MSC medium, passage after 85\% confluence. 0.5\% of trypsin and 0.2\% ethylenediaminetetraacetic acid (EDTA) can then be used to treat the cells for 5 minutes, wash, sediment at room temperature for 10 minutes (300 g), and plate in MSC medium at 5 x 10\textsuperscript{3} cells per cm\textsuperscript{2}. In the third passage, homogeneous MSC culture should be obtained.

Specific differentiation of hMSC can be initiated with commercially available hMSC differentiation kit.

OV (ICOVIR-5) can be added to the trypsinized, washed, and resuspended homogenous MSC culture in DMEM and incubated for 2 hours at 37 \textdegree C. The cells should then be washed and resuspended in DMEM 10\% FBS. EGFP-positive cells can be determined with flow cytometry to verify the transduction efficiency. After the infection, the cells will need to undergo a replication cycle to allow a high load of MSCs and OVs at around 80\%-100\% infection. Once validated, the cells should be washed twice with PBS and incubated with fresh medium. The harvesting and cell extracting process of OV carriers can be prepared by freeze-thawing.

\section{Expected outcomes}
According to all theories discussed in the part of \nameref{sec:discussion}, this method should be able to facilitate systemic administration of OVs by using the MSCs as a carrier to protect them from the immune system and directly transfer OVs to the cancer site where OVs are unloaded and directly infect cancer cells and cause oncolysis. If the infection is sustained long enough, continuous distress of cancer cells mediated by oncolysis will trigger the secretion of DAMPs recruiting the adaptive immune system to the site, thus inducing an anti-tumor response by the immune system. 

\begin{figure}[H]
\includegraphics[width=\linewidth]{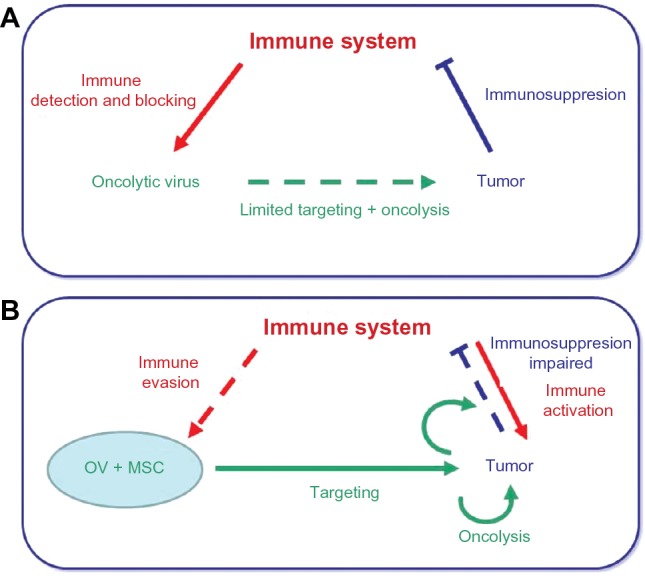}
\caption{MSCs used as carriers for systemically administered OVs. \textbf{(A)} The immune system directly attacks OVs, impeding them from reaching their targets. \textbf{(B)} MSCs shield OVs from the immune system and migrate them directly to the tumor, triggering direct oncolysis and an anti-tumor reaction mediated by the immune system.
Note: From "Patient-derived mesenchymal stem cells as delivery vehicles for oncolytic virotherapy: novel state-of-the-art technology," by \cite{RN87}, \textit{Oncolytic Virother}, 4, 149-155. Copyright 2015 by Dove Medical Press Limited under CC BY-NC license. Reprinted under license terms.}
\label{fig8}
\end{figure}

The mechanism of infection is similar to that of a local administration of OVs but uses MSCs as a carrier. This also eliminates the risk of deleterious effects associated with the systemic delivery of OVs \hyperref[fig8]{(Figure 8)}.

OVT seems likely to succeed current cancer therapies such as radiation and chemotherapy by nearly eliminating all side effects associated with radiotherapy and chemotherapy while also utilizing all of their advantages, thus paving the way for the next stage in cancer treatment. 

\FloatBarrier
\FloatBarrier

\section{Implications}
OVT will be one of the next therapy procedures that are both successful and safe while also being capable of curing all sorts of tumors, including the once incurable GBM. The National Institutes of Health has already begun phase one testing of the treatment, which has been effectively used to treat a variety of other malignant tumors, including liver and breast cancer. Clearly, this is a significant triumph for the millions of individuals who are now battling cancer, as well as for the loved ones of cancer patients worldwide.

Cancer has undoubtedly had a profound effect on society. Due to this exceptional tumor therapy and the tireless efforts of medical experts and scientists, cancer treatment has taken a quantum leap forward. Due to the simplicity and superiority of the technique compared to modern cancer therapies such as immunotherapy, chemotherapy, and radiation, it will be economical yet effective for a large number of patients battling cancer.  This is particularly remarkable as it represents a significant breakthrough in the treatment of late-stage cancers for which conventional cancer treatments have been ineffective.

I anticipate widespread adoption of OVT in medical oncology, which has enormous potential for the treatment of cancers in humans. I strongly believe that in the next five to ten years, OVT will be routinely used for central nervous system cancer, esophageal and lung tumors. 

\section{Data availability}

Data sharing not applicable to this report as no datasets were generated or analysed during the current study.

\section{Funding}
This article received no specific grant from any funding agency in the public, commercial, or not-for-profit sectors.
\section{Conflicts of interest}
The author reports no conflict of interest.

\bibliography{bibliography}

\end{document}